\documentclass[twocolumn,english,aps,superscriptaddress, pra,groupedaddress]{revtex4-2}

\usepackage[latin9]{inputenc}
\usepackage{amsmath}
\usepackage{amssymb}
\usepackage{graphicx}
\usepackage{babel}
\usepackage{mathrsfs}
\usepackage{amsfonts}
\usepackage{epstopdf}
\usepackage{caption}
\usepackage{multirow}

\usepackage{color}
\usepackage[normalem]{ulem}

\DeclareMathOperator{\Li}{Li}

\begin{document}
\title{Transport in p-wave interacting Fermi gases}

\author{Jeff Maki}
\affiliation{Department of Physics and HKU-UCAS Joint Institute for Theoretical and Computational Physics, The University of Hong Kong, Hong Kong, China}
\affiliation{Pitaevskii BEC Center, CNR-INO and Dipartimento di Fisica, Universit\`{a} di Trento, I-38123 Trento, Italy}
\author{Tilman Enss}
\affiliation{Institut f\"ur Theoretische Physik, Universit\"at Heidelberg, 69120 Heidelberg, Germany}

\begin{abstract}
The scattering properties of spin-polarized Fermi gases are dominated by p-wave interactions. Besides their inherent angular dependence, these interactions differ from their s-wave counterparts as they also require the presence of a finite effective range in order to understand the low-energy properties of the system. In this article we examine how the shear viscosity and thermal conductivity of a three-dimensional spin-polarized Fermi gas in the normal phase depend on the effective range and the scattering volume in both the weakly and strongly interacting limits. We show that although the shear viscosity and thermal conductivity both explicitly depend on the effective range near resonance, the Prandtl number which parametrizes the ratio of momentum to thermal diffusivity does not have an explicit interaction dependence both at resonance and for weak interactions in the low-energy limit. In contrast to s-wave systems, p-wave scattering exhibits an additional resonance at weak attraction from a quasi-bound state at positive energies, which leads to a pronounced dip in the shear viscosity at specific temperatures.
\end{abstract}

\date{\today}
\maketitle

\section{Introduction}

Spin-polarized Fermi gases have become an excellent platform for studying quantum systems with higher partial-wave interactions. Due to the Pauli exclusion principle, the leading short-ranged interactions are p-wave in nature ($\ell = 1$, where $\ell$ is the angular momentum quantum number). For this reason, there have been numerous theoretical investigations into the physics of p-wave Fermi gases, such as p-wave superconductivity and topological physics in two-dimensions \cite{Leggett75, Volovik88, Read00, Jiang18, Yang20, Gurarie07}, the three-body loss rate \cite{Suno03, Jona08, Schmidt20, Zhu22}, and its application to one-dimensional physics \cite{Cui16, Zhang17, Pan18, Yin18, Maki21, Sekino18, Sekino21}. Such physics is not purely theoretical, but has become the focus of recent experimental investigations both in $^{40}$K and in $^6$Li, as there is a p-wave Feshbach resonance \cite{Regal03,  Zhang04,Schunck05, Gaebler07, Fuchs08, Yoshida18, Chang20, Marcum20, Ahmed21, Jackson22, Chevy05, Inada08}.

The main difference between p-wave ($\ell = 1)$ and s-wave ($\ell=0$) scattering in 3D is the presence of the centrifugal barrier. The centrifugal barrier limits the overlap between two-body bound states and scattering states to a region at short inter-particle distances, making the resonance inherently narrow. As a corollary, the two-body bound state is quite long-lived near resonance, even for positive energies when the two-body bound state is inside the scattering continuum, i.e., the bound state becomes a long lived quasi-bound state. In other words, the wavefunction for the two-body bound state remains localized even near resonance, in contrast to s-wave systems where the two-body bound state size approaches infinity as one approaches resonance. 

The presence of the centrifugal barrier and the narrowness of the resonance are related to the relevancy of the effective range in the scattering amplitude. For p-wave scattering in 3D, the inverse p-wave scattering amplitude is:
\begin{equation}
f_{\ell =1}^{-1} = -i - \frac{1}{p^3 v} - \frac{R}{p} + O(p)
\label{eq:scattering_amplitude}
\end{equation}
where $p$ is the magnitude of the relative momentum, related to the relative scattering energy, $E = p^2/m$ with $m$ the single-particle mass. We also define $v$ as the 3D p-wave scattering volume with units of volume, and $R$ as the p-wave effective range parameter with units of momentum. The first term represents the unitary scattering, while the second and third terms define the scattering parameters.  As one can see from the effective range expansion, both the scattering volume and effective range terms in the scattering amplitude are parametrically 
more important than the unitary term at low energies. Furthermore, one can show that the low-energy limit, $E \ll R^2/m$, and the zero-range limit $R \to 0$, can not be taken simultaneously, in contradistinction to s-wave interactions. Thus the low-energy scattering physics depends on both the scattering volume and the effective range.

Since the effective range is a relevant quantity to understand the low-energy scattering, the energetics and dynamics will also crucially depend on the effective range. Previous studies have examined this in the context of the necessity of two thermodynamic contacts in describing the energetics in the normal phase \cite{Yu15, Yao18}, the Landau liquid parameters \cite{Ding19}, and the three-body recombination rate \cite{Jona08,Zhu22}. In these studies, the effective range was found to be important describing the  leading behaviour. This ought to be compared to s-wave physics where the effective range merely adds a perturbative correction \cite{Schwenk05, Lacroix16, Schonenberg17, Miller18}. 

In terms of transport, a previous study examined the bulk viscosity \cite{Maki20b} for the p-wave Fermi gas. There it was shown that the effective range produces a finite bulk viscosity at resonance (i.e., when $v^{-1}=0$) proportional to $\zeta \propto T^{5/2}/R^2$. In the weakly interacting limit, the bulk viscosity is proportional to $v^2T^{9/2}$. If one tries to take the zero-range limit, the bulk viscosity at resonance diverges, which is a hallmark of the relevancy of the effective range. This is in contrast to the s-wave case where the bulk viscosity only depends on the s-wave scattering length, $a$, in the strongly interacting limit, and vanishes at resonance, $a^{-1} = 0$ \cite{Enss19, Nishida19, Hofmann20, Dusling13, Fujii20, Fujii22}.

An interesting open question is how the remaining transport coefficients, the shear viscosity and thermal conductivity, depend on the scattering parameters in both the weakly and strongly interacting limit. For s-wave interactions, it is known that these two quantities become divergent near the non-interacting point ($a=0$) as $a^{-2}$, while near resonance the shear viscosity and thermal conductivity depend only on the equation of state, i.e., the density and the temperature \cite{Bruun05, Bruun07, Rupak07, Braby10, Enss11, Enss12, Bluhm17, Frank20, Hofmann20, Fujii21}. It is unclear how this picture is modified for p-wave Fermi gases in 3D, even in the experimentally applicable limit of a small but finite effective range. 

In this work we consider this issue and evaluate the shear viscosity, $\eta$, and thermal conductivity, $\kappa$, using the kinetic theory approach \cite{Landau, Smith}. We find that the shear viscosity and thermal conductivity scattering times explicitly depend on the scattering parameters, even at resonance. Although the shear viscosity and thermal conductivity explicitly depend on the interaction parameters, we show that the Prandtl number, which describes the ratio of momentum and thermal diffusion, approaches a universal constant in these two limits that does not explicitly depend on the interaction, similar to the case of s-wave physics.

The remainder of this article is organized as follows. In Sec.~\ref{sec:2body} we present the two-body scattering properties for spin-polarized Fermi gases and obtain the two-body T-matrix. From there we give a brief overview of the kinetic theory approach and how it applies to the shear viscosity and thermal conductivity in Sec.~\ref{sec:kin_theory}. We then present the results for the shear and thermal scattering times in Sec.~\ref{sec:scattering_times} for arbitrary values of the scattering volume. For negative scattering volumes, when the bound state becomes a long-lived quasi-bound state we find non-monotonic behaviour for the transport properties which is further discussed in Sec.~\ref{sec:shear_visc}. From there we discuss the Prandtl number in Sec.~\ref{sec:Pr}, and finally conclude our discussions in Sec.~\ref{sec:conc}.

\section{Two-Body p-Wave Scattering}
\label{sec:2body}

In this article we consider a single-channel model for a spin-polarized Fermi gas with p-wave interactions:
\begin{align}
H &= \int d^3{\bf x} \frac{1}{2}\nabla_{\bf x} \psi^{\dagger}({\bf x})\nabla_{\bf x}\psi({\bf x}) \nonumber \\
&+ \int d^3{\bf x} \frac{g}{4} \psi^{\dagger}({\bf x}) \overleftrightarrow{\nabla}_{\bf x} \psi^{\dagger}({\bf x})\psi({\bf x}) \overleftrightarrow{\nabla}_{\bf x} \psi({\bf x})
\label{eq:H}
\end{align}
where $\psi^{(\dagger)}({\bf x})$ is the annihilation (creation) operator for spin-polarized Fermions, the bidirectional gradient is $\overleftrightarrow{\nabla}_{\bf x}=(\overleftarrow{\nabla}_{\bf x}-\overrightarrow{\nabla}_{\bf x})/2$, $g$ is the p-wave coupling constant, and we have set $\hbar$ and the atomic mass, $m$, to unity. 

As is custom, we renormalize this theory by examining the two-body scattering. Consider the two-body T-matrix between states with center-of-mass momentum ${\bf Q}$, relative momenta ${\bf p}$ and ${\bf q}$, $| {\bf Q}/2 \pm {\bf p \ (q)}\rangle$,  and a total complex frequency  $E$. In the presence of the many-body background the T-matrix has the form \cite{Yao18, Maki20b}:
\begin{align}
\langle {\bf \frac{Q}{2} \pm p} | &\mathcal{T} | {\bf \frac{Q}{2} \pm q}\rangle = {\bf p \cdot q} \, T({\bf Q}, E) \nonumber \\
T^{-1}({\bf Q}, E) &= \frac{1}{24\pi} \left[ \frac{1}{v} + \left(E-\frac{Q^2}{4}+2\mu\right) R \right. \nonumber \\
&\left. + \left(-E + \frac{Q^2}{4}-2\mu\right)^{3/2} + A_\text{mb}({\bf Q},E)\right]
\label{eq:T_matrix}
\end{align}
where $\mu$ is the chemical potential, and $A_\text{mb}({\bf Q}, E)$ is the contribution due to the many-body background which is reported in Appendix \ref{appendix:mb_T_matrix}. Such a term captures the effects of Fermi blocking in the intermediate scattering states on an equal footing with the Fermi factors in the collision integral (see below,  Eq.~(\ref{eq:colision_integral})).

The T-matrix in the first line of Eq.~\eqref{eq:T_matrix} splits into two pieces \cite{T_matrix_Note}. The first piece is the form factor of the p-wave interaction potential, ${\bf p \cdot q}$, see also Eq.~(\ref{eq:H}). The second piece, $T({\bf Q},E)$, describes the dependence of the scattering on the center-of-mass momentum, ${\bf Q}$, and total energy, $E$. Due to the presence of the many-body background, the scattering is no longer Galilean invariant, which is described by the nontrivial dependence on the center-of-mass momentum in $A_\text{mb}({\bf Q},E)$. 

Equation~\eqref{eq:T_matrix} is already renormalized, and both the scattering volume, $v$, and effective range, $R$, are defined in terms of the coupling constant $g$ and the ultraviolet cutoff in the theory, $\Lambda$:
\begin{align}
\frac{1}{24\pi v} &= \frac{1}{g} - \frac{\Lambda^3}{18 \pi^2}, &  \frac{R}{24\pi} &= \frac{\Lambda}{6\pi^2}.
\label{eq:renormalization_conditions}
\end{align}
This renormalization ensures that the p-wave scattering amplitude in the absence of the many-body background has the form shown in Eq.~(\ref{eq:scattering_amplitude}) since:
\begin{equation}
f_{\ell=1} =  - \frac{p^3 T({\bf Q}, E_{o.s})}{24\pi},
\label{eq:f_T}
\end{equation}
where now $E_{o.s} = \frac{Q^2}{4}+p^2 -2\mu + i \delta$ is the on-shell energy for two-particle scattering.

In this theory there are two two-body bound states at energies defined as the poles of Eq.~(\ref{eq:T_matrix}). The first is a shallow dimer with $E_b = -1/(vR)$. For positive values of the scattering volume, this is a true two-body bound state. Therefore we call the regime where $E_b <0$ the BEC side. For negative values of the scattering volume, $E_b>0$, the dimer is a long-lived quasi-bound state. We label this side as the BCS side. The second bound state is a deep dimer with energy $-R^2$. Both the shallow and deep dimer bound states are three-fold degenerate for $\ell = 1$. The state with energy $-R^2$ is actually unphysical as it possesses a negative norm \cite{Nishida12, Braaten12}. To avoid this issue, we will work in the low-energy limit where all energy scales $E$ satisfy: $E \ll R^2$, so that the presence of such an unphysical state is unimportant.  

We note that often a single-channel model is inadequate for describing p-wave scattering in terms of the microscopic parameters that parametrize the two-body interaction potential, for e.g. the van der Waals length. In this study we are only interested in the general properties of the transport in terms of the low-energy scattering parameters, $v$ and $R$. For our purposes, the single-channel model suffices after the proper renormalization of the T-matrix is taken into account, see Eq.~(\ref{eq:w}). However, for more in-depth knowledge of the individual p-wave scattering parameters, a two-channel model is required \cite{Yao18, Maki20b}.







\section{Kinetic Theory}
\label{sec:kin_theory}
In order to calculate the shear viscosity and thermal conductivity we employ the standard kinetic theory approach using the Boltzmann equation \cite{Landau, Smith}
\begin{equation}
\frac{\partial n_{\bf p}}{\partial t} + {\bf p} \cdot \nabla_{\bf x} n_{\bf p} = I\left[n_{\bf p}\right],
\label{eq:Boltzmann_equation}
\end{equation} 
where $n_{\bf p} = n_{\bf p}({\bf x}, t)$ is the local quasiparticle distribution function. The collision integral, $I\left[n_{\bf p}\right]$, is defined as
\begin{align}
I&\left[n_{\bf p}\right] = \int \frac{d^3{\bf q}}{(2\pi)^3} \int \frac{d^3{\bf q'}}{(2\pi)^3}\int \frac{d^3{\bf p'}}{(2\pi)^3} W({\bf p,q,p',q'}) \nonumber \\
&\times \left[(1-n_{\bf p})(1-n_{\bf q})n_{\bf p'} n_{\bf q'} -(1-n_{\bf p'})(1-n_{\bf p'})n_{\bf p} n_{\bf q}\right]. \nonumber \\
\label{eq:colision_integral}
\end{align}
Equation~\eqref{eq:colision_integral} depends on the transition rate $W({\bf p,q,p',q'})$ between particles with incoming momenta $\bf p,q$ and outgoing momenta $\bf p',q'$, defined in terms of the on-shell T-matrix, $T({\bf Q,p}) = T({\bf Q}, E_{o.s.})$:
\begin{align}
W&({\bf p,q,p',q'}) = (2\pi)^4 \delta({\bf p + q- p'-q'}) \nonumber \\
& \times \delta\left(\frac{p^2}{2} + \frac{q^2}{2} - \frac{p'^2}{2}-\frac{q'^2}{2}\right)  \left( \frac{\bf p-q}{2} \cdot \frac{\bf p'-q'}{2}\right)^2  \nonumber \\
& \times \left| T\left({\bf p + q},  \frac{\bf p -q}{2} \right) \right|^2.
\label{eq:w}
\end{align}
Following kinetic theory we linearize the Boltzmann equation by writing $n_{\bf p} = n_{\bf p}^0 + \delta n_{\bf p}$, where $n_{\bf p}^0$ is the local equilibrium distribution function,
\begin{equation}
n_{\bf p}^0 = \left[e^{\beta \left(({\bf p - v})^2/2-\mu\right)} +1 \right]^{-1},
\end{equation}
that depends on the local inverse temperature, $\beta({\bf x},t) = 1/T({\bf x },t)$, velocity, ${\bf v}({\bf x},t)$ and chemical potential, $\mu({\bf x},t)$. We have muted the spatial and temporal coordinates of the thermodynamic variables for simplicity. The correction to the distribution function in response to an external perturbation is denoted by $\delta n_{\bf p}$. It is subject to the constraints of conserved number, momentum and energy,
\begin{align}
0 = \int \frac{d^3 {\bf p}}{(2\pi)^3} \lbrace 1, {\bf p}, \frac{p^2}{2} \rbrace \ \delta n_{\bf p} .
\end{align}
It can be written in the form $\delta n_{\bf p} = \beta n_{\bf p}^0 (1-n_{\bf p}^0) \phi({\bf p})$ with
\begin{align}
\phi({\bf p}) &= -\left[ \phi_{ij}({\bf p-v}) \frac{V_{i,j}}{2} - {\boldsymbol \phi}({\bf p -v})  \cdot \nabla_{\bf x} \ln \beta \right]
\label{eq:phi}
\end{align}
and $V_{i,j} = \partial_i v_j + \partial_j v_i - 2/3 \delta_{i,j} \nabla_{\bf x} \cdot {\bf v}$.

Due to rotational invariance, we only need to consider perturbations of the form $\phi_{xy}({\bf p})$ (shear) and $\phi_{x}({\bf p})$ (heat current). Following the kinetic theory approach \cite{Landau, Fujii20} the linearized Boltzmann equation leads to equations that determine $\phi_{xy}({\bf p})$ and $\phi_{x}({\bf p})$:
\begin{align}
&p_x p_y = \int \frac{d^3{\bf q}}{(2\pi)^3} \int \frac{d^3{\bf p'}}{(2\pi)^3} \int \frac{d^3{\bf q'}}{(2\pi)^3} n_{\bf q}^0(1-n_{\bf q}^0) \nonumber \\
&W({\bf p,q, p',q'}) \left(\phi_{xy}({\bf p}) + \phi_{xy}({\bf q}) - \phi_{xy}({\bf p'}) - \phi_{xy}({\bf q'}) \right), \nonumber \\
\label{eq:phixy} \\
&p_x \left(\frac{p^2}{2}- w\right)  = \int \frac{d^3{\bf q}}{(2\pi)^3} \int \frac{d^3{\bf p'}}{(2\pi)^3} \int \frac{d^3{\bf q'}}{(2\pi)^3} n_{\bf q}^0(1-n_{\bf q}^0) \nonumber \\ 
&W({\bf p,q, p',q'}) \left(\phi_{x}({\bf p}) + \phi_{x}({\bf q}) - \phi_{x}({\bf p'}) - \phi_{x}({\bf q'}) \right). \nonumber \\
\label{eq:phix} 
\end{align} 
In Eq.~(\ref{eq:phix}), $w$ is the enthalpy per particle, $w = (\varepsilon + p)/n$, with $\varepsilon$ as the energy density, $p$ the pressure, and $n$ the density. The functions $\phi_{xy}({\bf p})$ and $\phi_x({\bf p})$ which solve Eqs.~(\ref{eq:phixy}-\ref{eq:phix}) are then related to the shear viscosity, $\eta$, and thermal conductivity, $\kappa$, respectively:
\begin{align}
\eta &= \beta \int \frac{d^3{\bf p}}{(2\pi)^3} n_{\bf p}^0(1-n_{\bf p}^0)p_x p_y \phi_{xy}({\bf p}) \label{eq:eta_def}, \\
\frac{\kappa}{\beta} &= \beta \int \frac{d^3{\bf p}} {(2\pi)^3} n_{\bf p}^0(1-n_{\bf p}^0)p_x  \left(\frac{p^2}{2} - w\right) \phi_{x}({\bf p}). \label{eq:kappa_def}
\end{align}
 
To solve Eqs.~(\ref{eq:phixy}-\ref{eq:kappa_def}), we expand the functions $\phi_{xy}({\bf p})$ and $\phi_x({\bf p})$ in terms of a set of orthogonal basis functions:
\begin{align}
\phi_{xy}({\bf p}) &= \beta \sum_j c_j^{\eta} U_j^{\eta}({\bf p}), \\
\phi_{x}({\bf p}) &= \beta \sum_j c_j^{\kappa} U_j^{\kappa}({\bf p}),
\end{align}
where $c_j^{(\eta, \kappa)}$ are expansion coefficients, and the first basis functions are $U_1^{\eta} = p_x p_y$ and $U_1^{\kappa} = p_x (p^2/2 - w)$. In terms of these modes, Eqs.~(\ref{eq:phixy}-\ref{eq:kappa_def}) are more conveniently expressed as
\begin{align}
\delta_{i,1}  &= \sum_j \mathcal{A}^{\left(\eta,\kappa\right)}_{i,j} c_j^{(\eta, \kappa)},  & \mathcal{A}^{\left(\eta,\kappa\right)}_{i,j} = \frac{\left(U_i^{\left(\eta,\kappa\right)}, \mathcal{L} U_j^{\left(\eta,\kappa\right)}\right)}{\left(U_1^{\left(\eta,\kappa\right)},U_1^{\left(\eta,\kappa\right)}\right)}. \label{eq:boltzmann_equation_matrix}
\end{align}
In Eq.~(\ref{eq:boltzmann_equation_matrix}) we have introduced the inner product
\begin{align}
  \label{eq:inner}
\left(A,B\right) = \int \frac{d^3 {\bf p}}{(2\pi)^3} n_{\bf p}^0(1-n_{\bf p}^0) A({\bf p}) B({\bf p})
\end{align}
as well as the linearized collision integral operator
\begin{align}
\mathcal{L}B &= \frac{\beta}{1-n_{\bf p}^0}\int \frac{d^3 {\bf q}}{(2\pi)^3}\int \frac{d^3 {\bf p'}}{(2\pi)^3}\int \frac{d^3 {\bf q'}}{(2\pi)^3} n_{\bf q}^0(1-n_{\bf q}^0) \nonumber \\
& W({\bf p,q,p',q'}) \left(B({\bf p}) + B({\bf q}) -B({\bf p'}) - B({\bf q'})\right).
\end{align}
In this notation the shear viscosity and thermal conductivity have a simple form:
\begin{align}
\eta &=  \beta^2 \left(\mathcal{A}^{\eta}\right)^{-1}_{1,1}  \left(U_1^{\eta},U_1^{\eta}\right), \nonumber \\
\frac{\kappa}{\beta} &=  \beta^2 \left(\mathcal{A}^{\kappa}\right)^{-1}_{1,1} \left(U_1^{\kappa},U_1^{\kappa}\right).
\label{eq:transport_final}
\end{align}
The quantity $\beta\left(\mathcal{A}^{\eta, \kappa}\right) ^{-1}_{1,1}$ has units of time and defines the shear scattering time, $\tau_{\eta}$, and the thermal scattering time, $\tau_{\kappa}$ respectively:
\begin{equation}
\tau_{\eta,\kappa} =  \beta \left(\mathcal{A}^{\eta, \kappa}\right) ^{-1}_{1,1}.
\label{def:scattering_times}
\end{equation}
In principle, as the number of basis modes increases, the accuracy of the calculation improves. The case of a single mode is special, and is equivalent to the relaxation time approximation (RTA) \cite{Landau, Smith}. For our purposes we will work with two basis modes as that provides a drastic improvement in the calculation. This is discussed further below. We label this as beyond the relaxation time approximation (BRTA).

\section{Shear and Thermal Scattering Times}
\label{sec:scattering_times}
The linearized Boltzmann equation \eqref{eq:boltzmann_equation_matrix} is amenable to numerical calculation of the shear and thermal scattering times, $\tau_{\eta}$ and $\tau_{\kappa}$ via Eq.~(\ref{def:scattering_times}). For our purposes we will consider two basis modes both on the BCS and on the BEC side. The details of the calculation are shown in Appendix \ref{appendix:derivation}, while we only discuss the results here. 

\begin{figure}
\includegraphics[scale = 0.6]{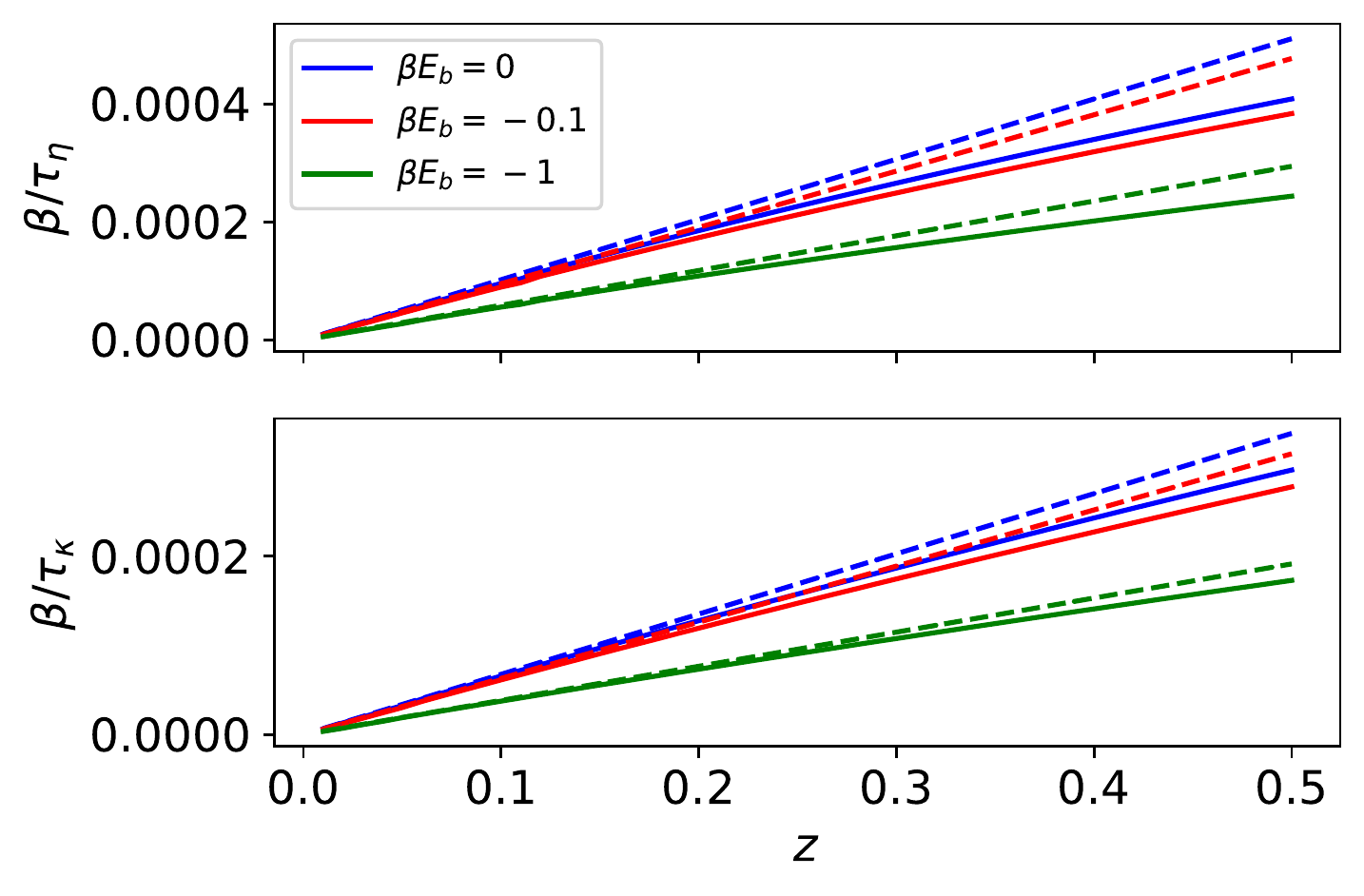}
\caption{Shear and thermal scattering rates as functions of fugacity, $z$. In all calculations we set the effective range $\beta^{1/2}R = 100$. The lines from bottom to top indicate the direction towards resonance, $|\beta E_b| \to 0$. The solid lines are the full numerical solution to the Boltzmann equation, while the dashed lines represent the corresponding high-temperature approximation which is linear in $z$.} 
\label{fig:scattering_times}
\end{figure}

The shear and thermal scattering times are presented in Fig.~\ref{fig:scattering_times} as functions of fugacity $z$ for various values of the binding energy $\beta E_b = -\beta/(vR)$. The dashed straight lines correspond to the high-temperature limit where $\beta /\tau_{\eta, \kappa} \propto z$. As the temperature is lowered (larger $z$), the deviations from the high-temperature limit become more pronounced, and we find that the shear and thermal scattering rates become smaller than the predictions from the high-temperature theory. This reduction of scattering is a consequence of Fermi blocking.

At resonance, $\beta E_b = 0$, the shear and thermal scattering times are proportional to $\beta R^2$ in the low-energy limit, up to a correction of order $O(1)$, as can be seen from Appendix \ref{appendix:derivation}. This is a consequence of defining the low-energy physics as: $\lbrace T, E,\mu,...\rbrace \ll R^2$. This leads us to an important result: the presence of the effective range is necessary for understanding the shear viscosity and the thermal conductivity in the strongly interacting limit. Such a conclusion is not necessarily obvious, as one might expect that for a small effective range the transport would predominantly depend on the equation of state since scale symmetry is only broken slightly. However, this is not the case as the low-energy constraint renders the effective range a relevant quantity. This should be contrasted to the spin-1/2 s-wave Fermi gas, where the resonant scattering times are only functions of the temperature and density because of the scale symmetry.

In the weakly interacting limit, a similar analysis shows that the shear and thermal scattering times are proportional to $v^{-2}$. This is the standard result for weakly interacting systems; the scattering times become infinite as one approaches the non-interacting limit. 

Although the shear and thermal scattering times have explicit interaction dependencies in the strongly and weakly interacting limits, the two scattering times actually have the same interaction dependencies to leading order. This is readily seen in Appendix \ref{appendix:derivation} where we provide explicit formulas in the high-temperature limit. This leads us to our second main result: since both scattering times have the same leading-order dependence on the scattering parameters both in the strongly and the weakly interacting limits, the ratio of the two scattering times in the strongly and weakly interacting limits is only a function of the equation of state:
$\tau_{\eta}/\tau_{\kappa} = \mathcal{F}\left(\beta\mu\right)+ O(1/(\beta R^2))$ with a dimensionless function $\mathcal{F}(x)$. Such behaviour also occurs for spin-1/2 Fermi gases with s-wave interactions, but this is because scale symmetry requires that the scattering times themselves are only functions of the equation of state. In the p-wave case, scale symmetry is still broken and thus the individual scattering times depend on the interactions, but their ratio will not depend on the scattering volume or effective range at leading order.

Let us first consider the results for the ratio of the shear to thermal scattering times in the high-temperature limit shown in Fig.~\ref{fig:BEC_BCS_Crossover} as a function of the bound state energy, $E_b$. In the RTA, the ratio of the scattering times is independent of the fugacity and the scattering parameters for arbitrary interaction strength, $\tau_{\eta}/\tau_{\kappa} = 2/3$, see Appendix \ref{appendix:derivation}. In the BRTA, the ratio of the scattering times becomes interaction dependent. In the strongly and weakly interacting limits, the ratios can be evaluated analytically to give:

\begin{figure}
\includegraphics[scale=0.55]{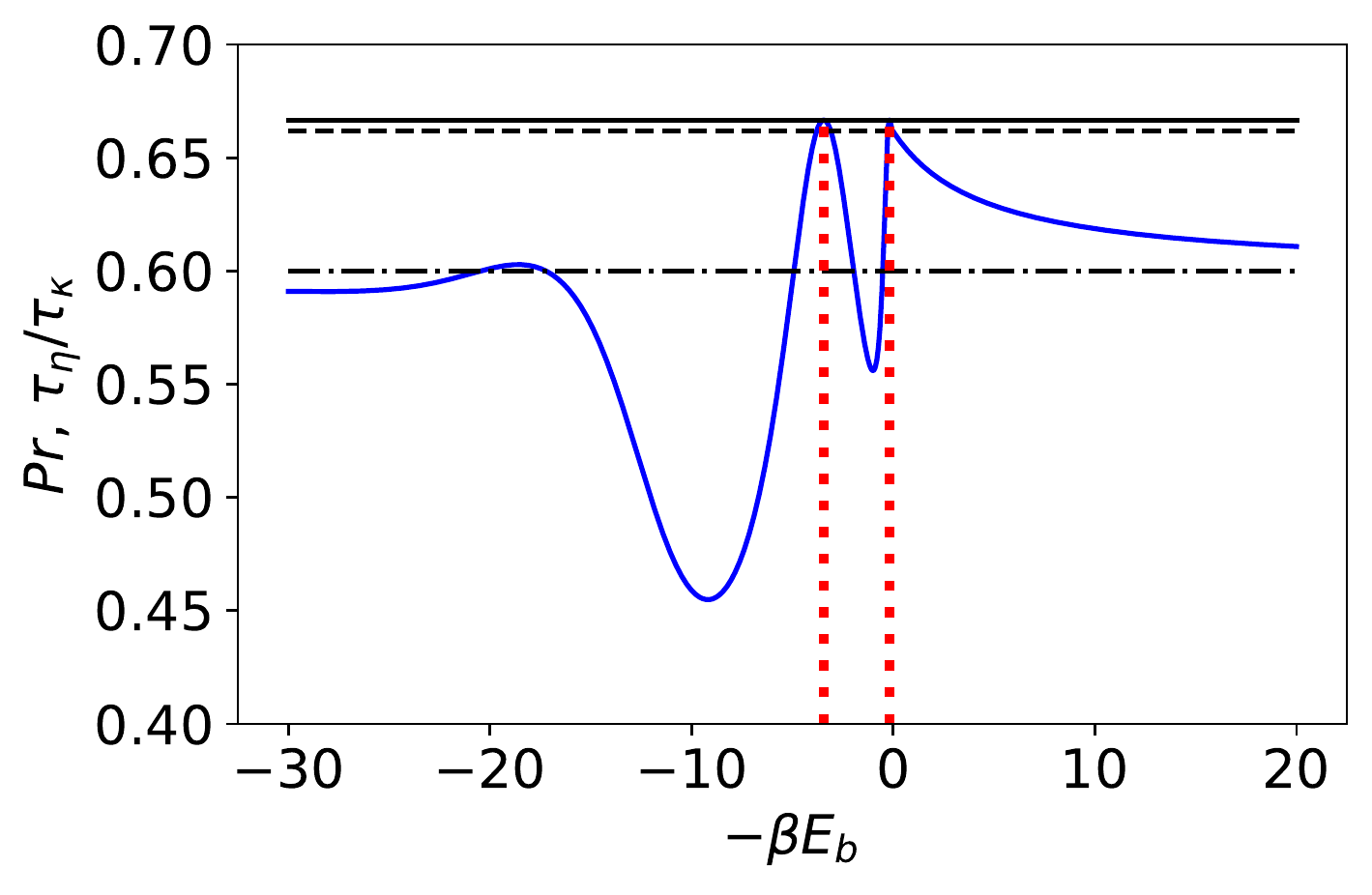}
\caption{BEC-BCS Crossover for the ratio of scattering times, $\tau_{\eta}/\tau_{\kappa}$ in the high-temperature limit with $\beta^{1/2} R = 100$. In this limit the ratio of scattering times coincides with the Prandtl number. The left hand side corresponds to the BCS side, while the right hand side is the BEC side. The solid, dashed, and dashed-dotted lines correspond to the RTA prediction, the BRTA prediction at resonance, and the BRTA prediction for weak interactions, respectively. The vertical dashed lines correspond to the points where the Prandtl number saturates the RTA value, $\beta E_b \approx 0, 7/2$.}
\label{fig:BEC_BCS_Crossover}
\end{figure}

\begin{align}
\left. \frac{\tau_{\eta}}{\tau_{\kappa}}\right|_{\textrm{res.}} &= \frac{2}{3}-\frac{14}{2727} \approx 0.662& \nonumber \\
\left. \frac{\tau_{\eta}}{\tau_{\kappa}}\right|_{\textrm{weak int.}} &=\frac{2}{3}-\frac{250}{3729} \approx 0.600
\label{eq:scattering_ratios}
\end{align}
In these limits the ratio of the scattering times does not explicitly depend on the interaction parameters, $v$ and $R$, the fugacity, $z$, and the inverse temperature, $\beta$. The individual values of $\tau_{\eta}$ and $\tau_{\kappa}$ are shown in Appendix \ref{appendix:derivation}.

On the BEC side (right) the behaviour of the ratio of scattering times is monotonic, while on the BCS side (left) there are two point where it saturates the RTA value. This can be understood as a consequence of the quasi-long-lived bound state, see Appendix \ref{appendix:derivation}. Specifically these resonances occur when the off-diagonal matrix elements of $\mathcal{A}^{(\eta,\kappa)}$ vanish. To leading order this occurs when the two-body bound state energy is approximately: $\beta E_b \approx 0, 7/2$. Since this resonance requires the two-body bound state energy to be positive, this secondary resonance where the ratio of scattering time saturates the RTA value can not occur for s-wave interacting systems. 

\begin{figure}
\includegraphics[scale=0.55]{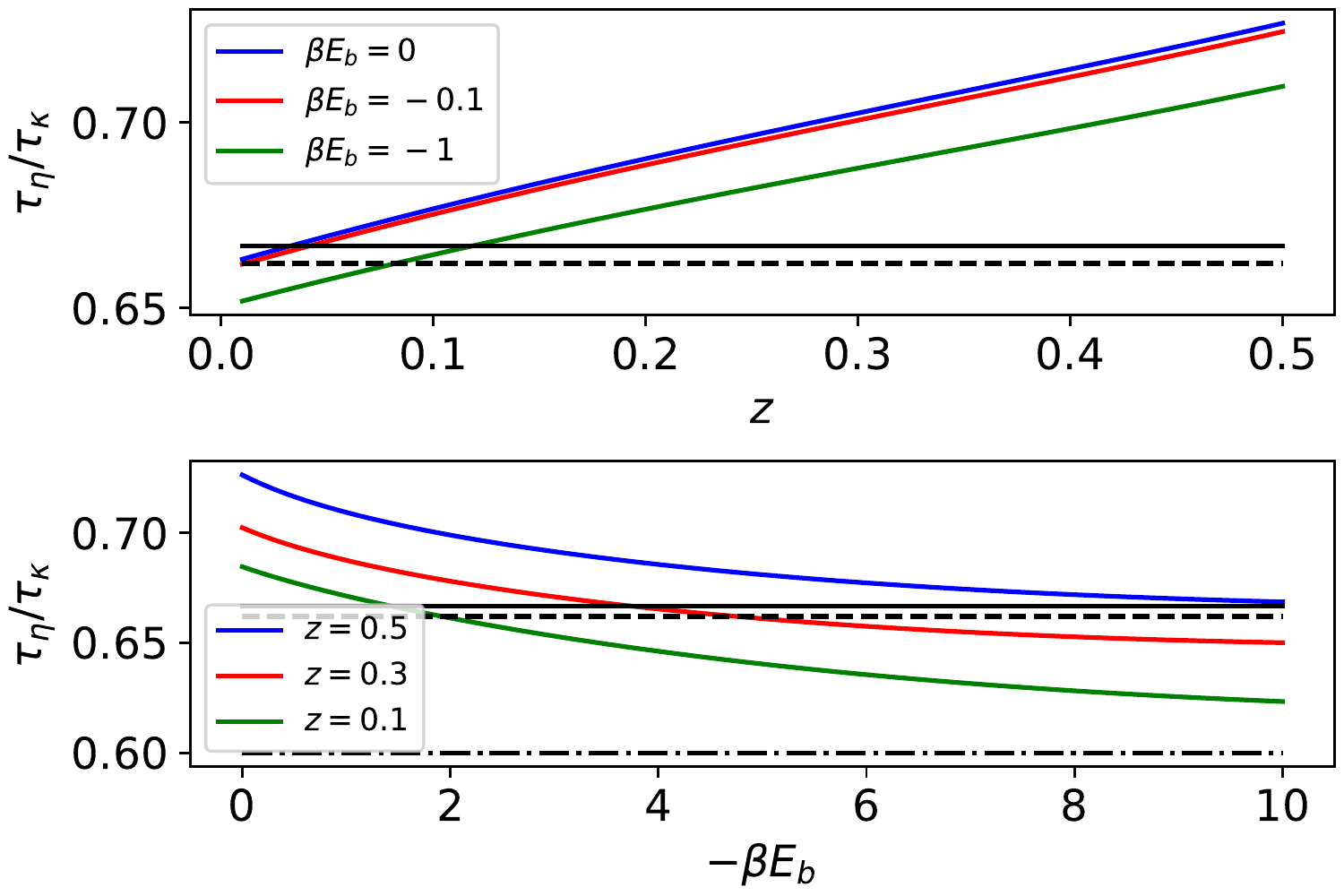}
\caption{Ratio of shear to thermal scattering times as a function of the fugacity $z$ (top), and the binding energy $E_b$ (bottom). The black solid line, dashed line, and dashed-dotted line correspond to the high-temperature results in the RTA, the resonant limit in the BRTA, and the weakly interacting limit in the BRTA, respectively. The lines from bottom to top indicate the direction towards: top) resonance, $|\beta E_b| \to 0$, and bottom) the low-temperature limit.}
\label{fig:ratios_BEC}
\end{figure}

As one lowers the temperature however, we do expect deviations to occur from the Fig.~\ref{fig:BEC_BCS_Crossover}. In particular, we expect that the ratio of the scattering times to leading order in $1/(\beta R^2)$ will be a dimensionless function of the equation of state, but not of the interaction strength, in the strongly and weakly interacting limit. In Fig.~\ref{fig:ratios_BEC} we examine the ratio of the two scattering times both a) as a function of fugacity and b) as a function of binding energy $E_b$ on the BEC side. The universal limits in Eq.~\eqref{eq:scattering_ratios} are shown as the dashed and dash-dotted lines, respectively. The black solid line corresponds to the RTA result of $2/3$. The two major trends are an increase of the ratio of scattering times as the temperature is lowered, and a decrease in the ratio as one goes to the weakly interacting limit. On the BCS side, $v<0$, we find that the many-body corrections are not as important as the scattering physics is highly dominated by the quasi-long-lived bound state. Hence the physics is accurately captured by the high-temperature physics. Thus even in the presence of the many-body background, there will still be a non-monotonic behaviour of the ratio of the scattering times, with the BRTA saturating to the RTA value at around $\beta E_b \approx 7/2$.

\section{Minimum in the Shear Viscosity and Thermal Conductivity on the BCS Side}
\label{sec:shear_visc}

The presence of the quasi-bound state and the non-monotonic behaviour of the ratio of the scattering times on the BCS side have important consequences for the shear viscosity and thermal conductivity. In particular, if the average energy of the atoms, which is proportional to $T$, is of the order of the quasi-bound state energy, the scattering will become maximal and equivalently the transport coefficient will exhibit a minimum. Thus it is instructive to look at the shear viscosity at fixed density and variable temperature, as we expect a dip to occur on the BCS side when the temperature is comparable to the quasi-bound state energy.

\begin{figure}
\includegraphics[scale=0.6]{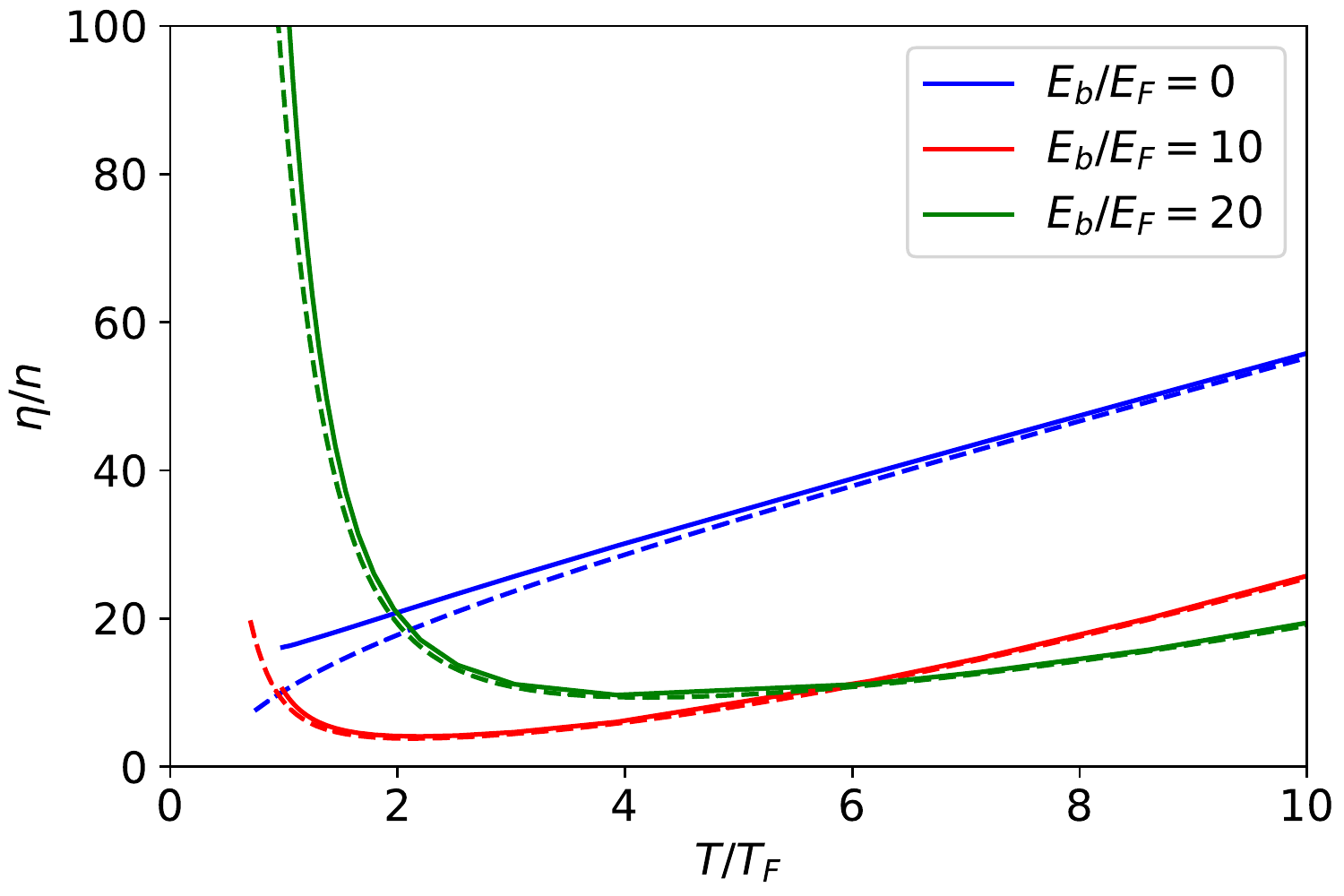}
\caption{Shear viscosity as a function of temperature for fixed density and various values of the binding energy, $E_b$, on the BCS side ($E_b >0$). In this figure, $T_F$ and $E_F$ are the Fermi temperature and energy, while the solid and dashed lines correspond to the RTA calculation and the high-temperature limit respectively. The dip in the shear viscosity is associated with an increased scattering at the quasi-bound state energy. This dip in the shear viscosity does not appear at resonance, $E_b = 0$, as zero-energy scattering is strongly suppressed for p-wave interactions. Similar physics will occur for the thermal conductivity.}
\label{fig:dip}
\end{figure}

We investigated this issue by calculating the shear viscosity at fixed density as a function of temperature within the RTA. Corrections from the BRTA are smaller than the Fermi blocking effect included in the RTA at lower temperatures and do not produce qualitative differences, see Fig.~\ref{fig:brta_to_rta} of Appendix~\ref{appendix:derivation}. In order to address the thermodynamics, we have assumed a noninteracting equation of state of spin-polarized Fermions. Such an approximation is reasonably valid in the BCS limit where no molecules exist, in contrast to the BEC limit discussed below where a Bose-Fermi model is used \cite{Yao18}. The results of this calculation are shown in Fig.~\ref{fig:dip} for $E_b = 20 E_F$, $E_b = 10 E_F$ and $E_b = 0$ (resonance).

As one can see from Fig.~\ref{fig:dip}, there is a dip in the shear viscosity due to the resonant scattering at the quasi-bound state energy. The minimum in the shear viscosity occurs roughly for $T/T_F \approx 0.2 E_b/E_F$. At resonance, there is no quasi-bound state and hence there is no minimum of the shear viscosity in the normal state. It is interesting to note that for p-wave scattering this minimum in the shear viscosity appears for $T>T_F$, while for s-wave Fermi gases it occurs at lower temperatures \cite{Enss11}.  Similarly, when the shear viscosity is expressed in units of entropy density $s$, it will also have a minimum. For example when $E_b/E_F=5$,  $\eta/s$ reaches a minimum value of approximately  $0.5\hbar/k_B$ at a temperature of $T\approx T_F$, comparable to the value found for the s-wave unitary Fermi gas \cite{Enss11} and slightly larger than the Kovtun-Son-Starinets bound \cite{Kovtun05}.

\section{Prandtl Number}
\label{sec:Pr}

Given the ratio of the shear to thermal scattering times, it is straightforward to calculate the Prandtl number. The Prandtl number is the ratio of the shear to thermal diffusivity, defined as
\begin{equation}
\Pr = \frac{\eta}{\kappa} \frac{C_p}{n},
\end{equation}
where $C_p$ is the specific heat at constant pressure per unit volume, and $n$ is the density. In this work we calculate the shear viscosity and thermal conductivity according to Eq.~(\ref{eq:transport_final}), while we calculate the specific heat at constant pressure and density using the Bose-Fermi model, see below.

At high temperatures, $C_p/n = 5/2$ while $\eta/\kappa = 2\tau_{\eta}/5\tau_{\kappa}$, so that $\Pr = \tau_{\eta}/\tau_{\kappa}$. This was presented in Fig.~\ref{fig:BEC_BCS_Crossover} for the whole BEC-BCS crossover. As one can see, the Prandtl number is a nonmonotonic function of $E_b$ on the BCS side, while it is monotonic on the BEC side.

We evaluate the thermodynamics using the Nozieres-Schmitt-Rink (NSR) approximation \cite{Nozieres1985}. This scheme involves calculating the free energy using the in-medium T-matrix. In this way we calculate the free energy and the scattering times consistently at the same level of approximation. Such a calculation was done previously in the context of p-wave Fermi gases using a two-channel model \cite{Yao18}, but we have verified that their predictions are equivalent to the single-channel model after renormalization of the many-body T-matrix, see Eq.~(\ref{eq:renormalization_conditions}).

\begin{figure}
\includegraphics[scale=0.55]{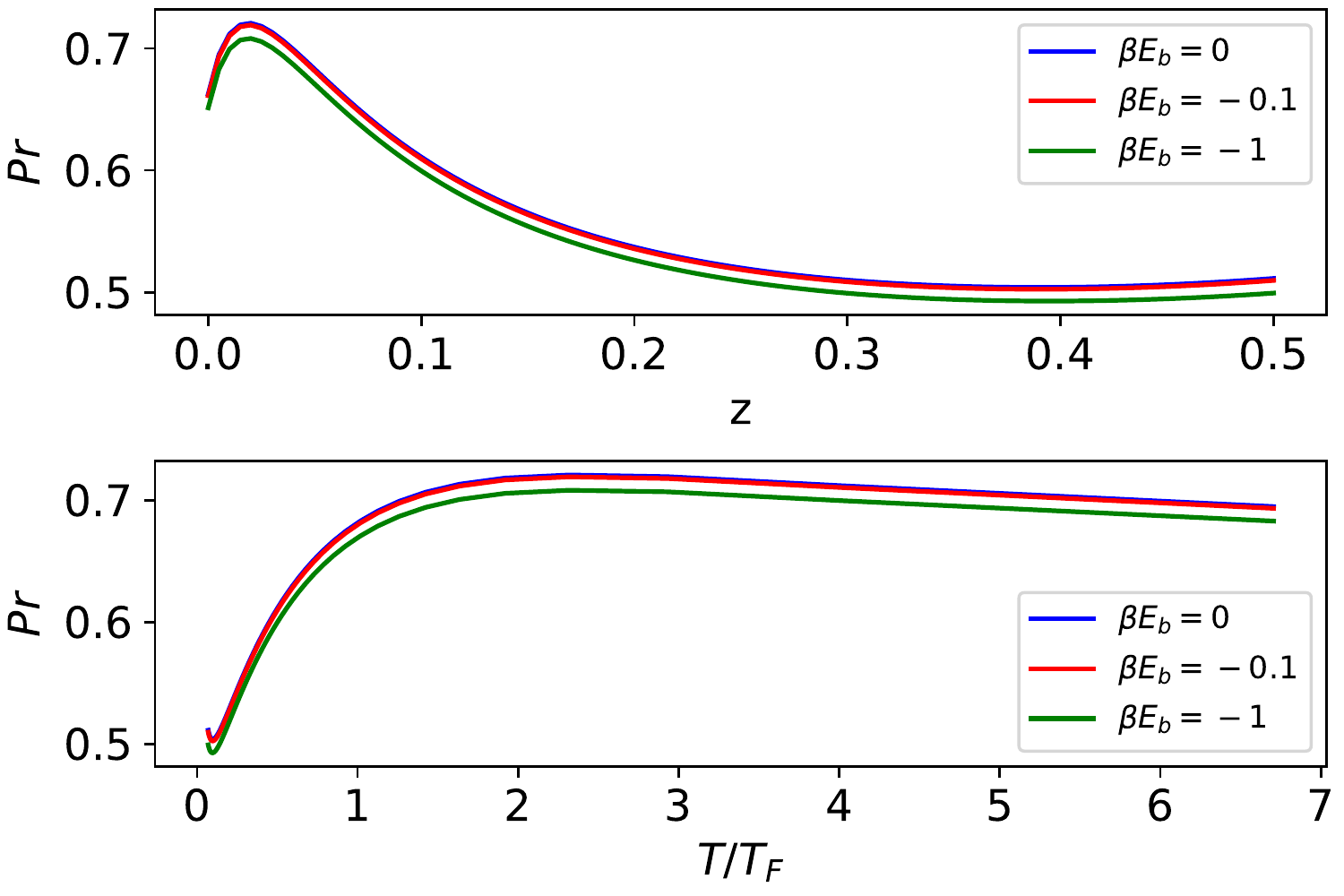}
\caption{Prandtl number on the BEC side as a function of fugacity, $z$, and density through the Fermi temperature $T_F$, for various values of the binding energy, $\beta E_b$. For the top (bottom) figures, the lines from bottom to top denote towards (away from) resonance. In general, decreasing the interaction decreases the Prandtl number, while lowering the temperature (increasing the fugacity or increasing the density) tends to an ultimate decrease of the Prandtl number by about $25\%$.}
\label{fig:PR}
\end{figure}

The free-energy calculated in the NSR scheme is equivalent to  a model describing a non-interacting mixture of bosons and fermions (see Appendix~\ref{appendix:Bose_Fermi}). The Bose-Fermi model is an accurate description of the thermodynamics of a normal state p-wave Fermi gas in 3D for the entire BEC-BCS crossover \cite{Yao18}. Here we restrict ourselves to the high-temperature limit which is defined as: $E_b \ll T \ll R^2$.  The results of our calculations are shown in Fig.~\ref{fig:PR}. As one can see for the BEC side, the Prandtl number is non-monotonic, but ultimately decreases as temperature decreases or density increases to about $25\%$ of the resonant value. At fixed fugacity, the Prandtl number also decreases slightly as the interaction parameters are decreased.

\section{Conclusions}
\label{sec:conc}

In this article we have examined the shear viscosity and thermal conductivity of a 3D p-wave spin-polarized Fermi gas. We found that the scattering times are proportional to $R^2$ near resonance and $v^{-2}$ for weakly interacting systems. This means that the transport properties explicitly depend on the scattering parameters for arbitrary interaction strengths. However, the Prandtl number in the weakly and strongly interacting limits is indeed a function of only the equation of state, like the Prandtl number for spin-1/2 s-wave Fermi gases. Unlike the spin-1/2 s-wave Fermi gas, there is no scale symmetry restricting the Prandtl number to be this way.

Our analysis is valid at high temperatures in the normal phase. It quantifies the role of two-body correlations in the transport properties of spin-polarized Fermi gases in 3D. As one goes to lower temperatures, it is important to also account for three-body correlations and losses. Such losses have been previously examined \cite{Suno03, Jona08, Schmidt20, Zhu22, Regal03, Zhang04, Chevy05, Gaebler07, Fuchs08, Inada08}, and were found to become quite strong at the p-wave Feshbach resonance. These three-body losses are suppressed in the high-temperature limit as they are of order $z^3$. Obviously these losses will have an important contribution to the transport at lower temperatures, but for temperatures $T \gtrsim T_F$ the transport properties are determined by the two-body scattering processes discussed in this work. We predict a pronounced dip in the viscosity at temperatures above $T_F$ that could be observed in experiment.

\section{Acknowledgements}
This work is supported by the Deutsche Forschungsgemeinschaft (DFG, German Research Foundation), project-ID 273811115 (SFB1225 ISOQUANT) and under Germany's Excellence Strategy EXC2181/1-390900948 (the Heidelberg STRUCTURES Excellence Cluster), as well as HK GRF 17304820, 17304719, 17305218 and CRF C6009-20G and C7012-21G. The authors would like to thank Shizhong Zhang for useful discussions.

\appendix

\numberwithin{equation}{section}
\renewcommand\theequation{\Alph{section}.\arabic{equation}}

\section{Many-Body Contribution to the Two-Body T-matrix}
\label{appendix:mb_T_matrix}

The many-body contribution to the two-body T-matrix has the form:
\begin{align}
A_\text{mb}(Q,&E) =  24\pi \nonumber \\
& \times \int \frac{d^3{\bf k}}{(2\pi)^3} k^2 \frac{1-u^2}{2} \frac{n_F(\xi_k)}{E -\frac{1}{2}Q^2 - k^2 - Q k u + 2\mu} \nonumber \\
\label{app:Amb_1}
\end{align}
where $u = \hat{Q} \cdot \hat{k}$, $\xi_k = k^2/2 - \mu$, and $n_F(x) = (e^{\beta x}+1)^{-1}$ is the Fermi-Dirac distribution function. In writing Eq.~(\ref{app:Amb_1}), we have neglected terms that produce a contribution to the T-matrix of the form: $\left({\bf p \cdot Q}\right) \left({\bf Q \cdot q} \right)\tilde{T}(E)$ as these vanish when the ultraviolet cutoff is taken to infinity.

The angular integration in Eq.~(\ref{app:Amb_1}) can be performed analytically to give:
\begin{align}
A_\text{mb}(Q,E) &=  \int_0^{\infty} \frac{dk}{2\pi^2} \ k^4 \frac{e^{-\frac{\beta}{2}k^2}z}{1+e^{-\frac{\beta}{2}k^2}z} \nonumber \\
&\left[ \frac{E + 2\mu - k^2 - Q^2/2}{2 Q^2 k^2} \right. \nonumber \\
&\left. + \frac{(E + 2\mu - k^2 - Q^2/2)^2 - Q^2 k^2}{4Q^3 k^3} \right. \nonumber \\
&\left. \times \ln\left[\frac{E + 2\mu - \frac{1}{2}k^2 - \frac{1}{2}(Q+k)^2}{E + 2\mu - \frac{1}{2}k^2 - \frac{1}{2}(Q-k)^2}\right]\right].
\label{app:Amb_2}
\end{align}
For our purposes we evaluate Eq.~(\ref{app:Amb_2}) numerically.

\section{Derivation of the Shear and  Thermal Matrices in Eq.~(\ref{eq:boltzmann_equation_matrix})}
\label{appendix:derivation}

In this appendix we derive the relevant expressions for the matrices $\mathcal{A}^{(\eta, \kappa)}$ in Eq.~(\ref{eq:boltzmann_equation_matrix}), using two basis modes. For both the shear viscosity and thermal conductivity the matrices $\mathcal{A}^{(\eta, \kappa)}$ have the form 
\begin{align}
\mathcal{A}_{i,j}&= \frac{\beta}{\left(U_1,U_1\right)}\int_0^{\infty} \frac{dQ}{2\pi^2} \int_0^{\infty} \frac{dp}{2\pi^2}  \int \frac{d\Omega_Q}{4\pi}\int \frac{d\Omega_p}{4\pi}\int \frac{d\Omega_q}{4\pi}  \nonumber\\
& \frac{Q^2 p^7}{4\pi}\left(\hat{\bf p}\cdot \hat{\bf q}\right)^2 |T({\bf Q,p})|^2\nonumber \\
&\times \frac{1}{2} \frac{1}{\cosh(a) + \cosh(b \hat{\bf Q}\cdot \hat{\bf p})}\frac{1}{2} \frac{1}{\cosh(a) + \cosh(b \hat{\bf Q}\cdot \hat{\bf q})} \nonumber \\
&\times\left[U_i({\bf \frac{Q}{2}+p}) + U_i({\bf \frac{Q}{2}-p})\right]\nonumber \\
&\left[U_j({\bf \frac{Q}{2}+p}) + U_j({\bf \frac{Q}{2}-p})- U_j({\bf \frac{Q}{2}+q}) - U_j({\bf \frac{Q}{2}-q})\right]
\label{app:boltzman_modes}
\end{align}
where we have used the identity
\begin{align}
n_{\bf \frac{Q}{2}+p} & n_{\bf \frac{Q}{2}-p}  (1- n_{\bf \frac{Q}{2}+q})(1- n_{\bf \frac{Q}{2}-q}) = \nonumber \\
&\frac{1}{4} \frac{1}{\cosh\left(a\right)  + \cosh\left(b\hat{\bf Q} \cdot \hat{\bf p}\right)} \frac{1}{\cosh\left(a\right)  + \cosh\left(b \hat{\bf Q} \cdot \hat{\bf q}\right)}
\label{eq:occupations}
\end{align}
as well as $a = \beta(Q^2/8  +p^2/2 - \mu)$ and $b =  \beta Q p/2$. We have also explicitly suppressed the $\eta$ and $\kappa$ indices in Eq.~(\ref{app:boltzman_modes}) as it is valid for both the shear viscosity and thermal conductivity.

Let us begin by considering the shear viscosity. The two basis modes we will consider are:
\begin{align}
U_{1}^{\eta} &= p_x p_y, & U_{2}^{\eta}&= p_x p_y 
\beta \left(p^2 - w_{\eta}\right).
\end{align} 

\noindent The constant $w_{\eta}$ can be determined from the Gram-Schmidt method, and is such that the two modes are orthogonal with inner product \eqref{eq:inner}. Given these two modes one can then evaluate Eq.~(\ref{app:boltzman_modes}).

However, special care is needed in performing the angular integrations. As an example consider the case $i=1$ and $j=1$. In this case Eq.~(\ref{app:boltzman_modes}) becomes
\begin{align}
\mathcal{A}^{\eta}_{1,1} &= \frac{\beta}{\left(U_1,U_1\right)}\int_0^{\infty} \frac{dQ}{2\pi^2} \int_0^{\infty} \frac{dp}{2\pi^2}  \int \frac{d\Omega_{\bf Q}}{4\pi} \nonumber\\
&\sum_{\mu,\nu= x,y,z}\frac{Q^2 p^{11}}{\pi} |T({\bf Q,p})|^2 \left[F_{\eta,2}^{\mu,\nu} F_{\eta,0}^{\mu,\nu} - F_{\eta,1}^{\mu,\nu} F_{\eta,1}^{\mu,\nu} \right] .
\label{app:A11}
\end{align}
The functions $F_{\eta,n}^{\mu,\nu}$ are the angular averages
\begin{align}
F_{\eta,n}^{\mu,\nu} &= \left\langle \frac{1}{2} \frac{1}{\cosh(a) + \cosh(b \cos{\theta_{\bf p}})} \hat{\bf p}_{\mu} \hat{\bf p}_{\nu} \left(\hat{\bf p}_x \hat{\bf p}_y\right)^n  \right\rangle_{\Omega_{\bf p}}
\label{app:F}
\end{align} 
where $a = \beta(Q^2/8+p^2/2 - \mu)$ and $b =  \beta Q p/2$, while $\langle \cdot \rangle_{\Omega_{\bf p}}$ denotes angular averages over $\hat{\bf p}$. As shown in Appendix \ref{appendix:angular_averages}, the angular averages can be decomposed into a sum of terms depending on different components of the center of mass unit vector, $\hat{\bf Q}$, and the integrals:
\begin{equation}
I_n(Q,p) = \int_{-1}^{1} \frac{dx}{4} \frac{P_n(x)}{\cosh(a) + \cosh(b x)}
\end{equation}
where $P_n(x)$ is the Legendre polynomial. 

\begin{figure}
\includegraphics[scale=0.6]{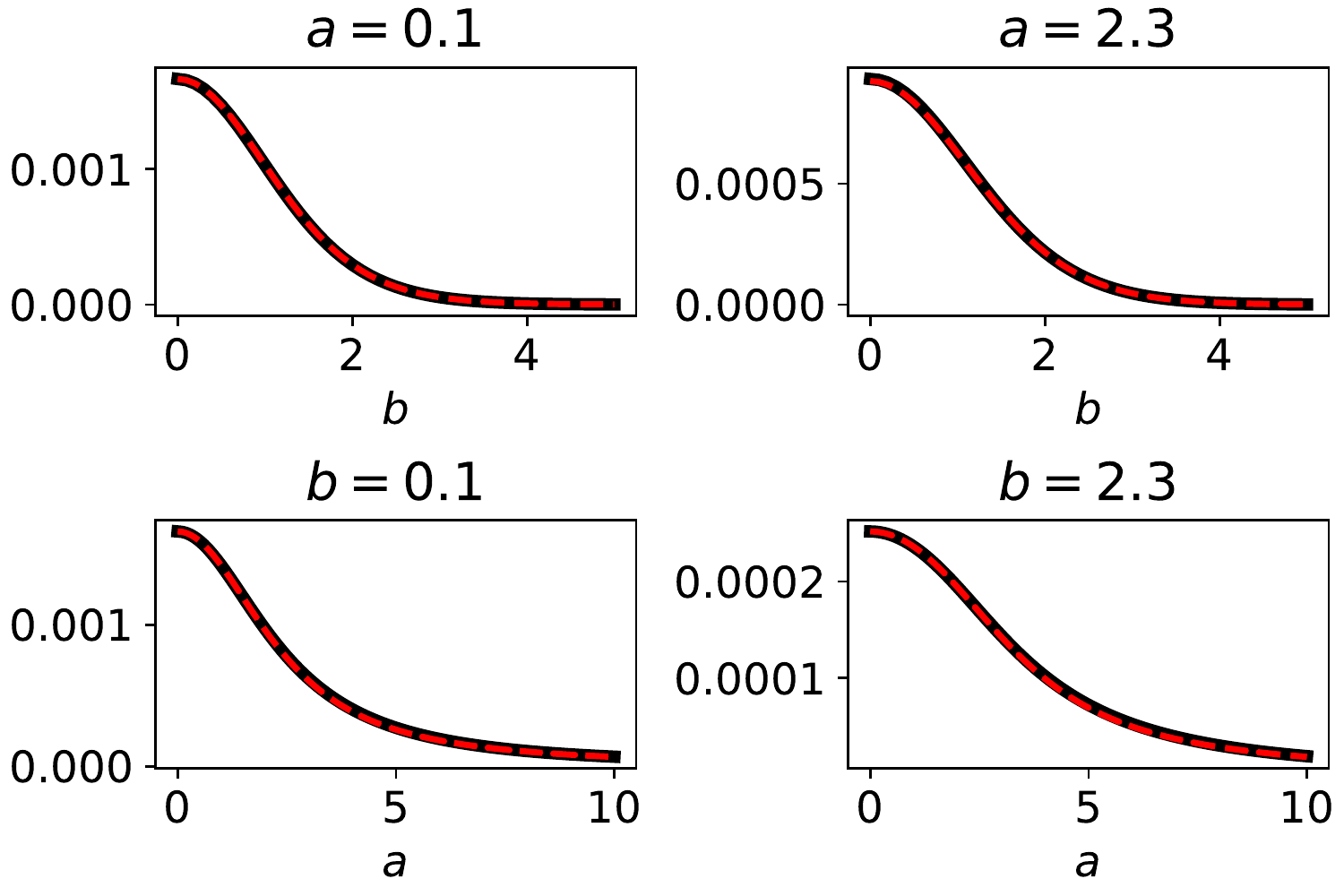}
\caption{Comparison of the full angular average of Eq.~(\ref{app:A11}) (black line) and the solution from Eq.~(\ref{app:approx_fn}) (red-dashed line) for various values of $a = \beta(Q^2/8  +p^2/2 - \mu)$ and $b =  \beta Q p/2$. The approximation is quite successful at capturing the full structure of the angular average for the desired parameter regime, $a,b \lesssim 1$.} 
\label{fig:ang}
\end{figure}

The dominant contribution to Eq.~(\ref{app:F}) comes from the term proportional to $I_0(Q,p)$, which is equivalent to making the substitution:
\begin{align}
F_{\eta,n}^{\mu,\nu} \approx I_0(Q,p) \left\langle \hat{\bf p}_{\mu} \hat{\bf p}_{\nu} \left(\hat{\bf p}_x \hat{\bf p}_y\right)^n  \right\rangle_{\Omega_{\bf p}}.
\label{app:approx_fn}
\end{align}
This approximation becomes exact in the high-temperature limit where Eq.~\eqref{eq:occupations} is angular independent and can be pulled out of the angular average, such that only the term proportional to $I_0$ survives. To test the accuracy of this approximation at lower temperatures we plotted the full angular average contained in  Eq.~(\ref{app:A11}) (solid black line) versus the result after the approximation in Eq.~(\ref{app:approx_fn}) (red-dashed line) for both fixed $a$ and $b$ in Fig.~\ref{fig:ang}. In the non-degenerate regime, the dominant contribution to $F_n^{\mu,\nu}$ comes from $a, b \lesssim 1$. The approximation of retaining only the terms proportional to $I_0$ is exceptionally good and simplifies the calculation enormously especially in the required parameter regime, cf.\ also Ref.~\cite{Enss12}.

With these approximations the $\mathcal{A}^{\eta}_{1,1}$ becomes
\begin{align}
\mathcal{A}^{\eta}_{1,1} &= \frac{\beta}{\left(U_1^\eta,U_1^\eta\right)}\int \frac{dQ}{2\pi^2} \int \frac{dp}{2\pi^2} \frac{Q^2p^{11}}{75\pi} \left|T({\bf Q,p})\right|^2 I_0^2(Q,p).
\end{align}
One can then repeat the following analysis for the other elements of the matrix $A^{\eta}$. The final result is:
\begin{align}
&\mathcal{A}^{\eta} = \frac{\beta}{\left(U_1^{\eta},U_1^{\eta}\right)} \nonumber \\
&\times\int_0^{\infty} \frac{dQ}{2\pi^2} \int_0^{\infty} \frac{dp}{2\pi^2} \frac{Q^2p^{11}}{75\pi} \left|T({\bf Q,p})\right|^2 I_0^2(Q,p) \nonumber\\
&\begin{bmatrix}
1 & \beta\left(\frac{7Q^2}{12}+p^2 - w_{\eta}\right)\\
\beta \left(\frac{7Q^2}{12}+p^2 - w_{\eta}\right) & \beta^2 \left(\left(\frac{7Q^2}{12}+p^2 - w_{\eta}\right)^2 + \frac{7Q^4}{90}\right)
\end{bmatrix}.
\label{app:shear_A}
\end{align}

An identical analysis can be done for the thermal conductivity. The two basis modes are given by
\begin{align}
U_1^{\kappa} &= p_x \left(\frac{p^2}{2} -w\right), & U_2^{\kappa} &= p_x \left(\frac{p^2}{2} -w\right)\beta \left(\frac{p^2}{2} - w_{\kappa}\right),
\end{align}
where $w$ is the enthalpy per particle and $w_{\kappa}$ is a constant that ensures that the two basis modes are orthogonal, similar to the case for shear viscosity. 

We again assume that the angular averages can be approximated in the manner of Eq.~(\ref{app:approx_fn}). The final result for the matrix $\mathcal{A}^{\kappa}$ is:
\begin{align}
&\mathcal{A}^{\kappa} = \frac{\beta}{\left(U_1^{\kappa},U_1^{\kappa}\right)} \nonumber \\
&\times\int_0^{\infty} \frac{dQ}{2\pi^2} \int_0^{\infty} \frac{dp}{2\pi^2} \frac{Q^4p^{11}}{270\pi} \left|T({\bf Q,p})\right|^2 I_0(Q,p)^2 \nonumber\\
&\begin{bmatrix}
1 & \beta(\frac{7Q^2}{20} +p^2 - w - w_{\kappa})\\
\beta(\frac{7Q^2}{20} +p^2 - w - w_{\kappa}) & \beta^2\bigl((\frac{7Q^2}{20} +p^2 - w - w_{\kappa})^2 + \frac{3Q^4}{200}\bigr)
\end{bmatrix}
\label{app:thermal_A}
\end{align}


In the high-temperature limit, one can analytically perform the integration over $Q$ to obtain:
\begin{align}
\mathcal{A}^{\eta} = \frac{2^{3/2}z}{75 \pi} &\int_0^{\infty} \frac{d\epsilon}{(2\pi)^2} e^{-\epsilon} \epsilon^5 |T(0,\sqrt{\epsilon})|^2 \nonumber \\
&\times \begin{bmatrix}
1 & \epsilon - \frac{7}{2} \\
\epsilon - \frac{7}{2} & \left(\epsilon -\frac{7}{2} \right)^2 +  \frac{77}{6} 
\end{bmatrix},
\label{app:highT_shear_A}\\
\mathcal{A}^{\kappa} = \frac{2}{3}\frac{2^{3/2}z}{75 \pi} &\int_0^{\infty}  \frac{d\epsilon}{(2\pi)^2} e^{-\epsilon} \epsilon^5 |T(0, \sqrt{\epsilon})|^2 \nonumber \\
&\times\begin{bmatrix}
1 & \epsilon - \frac{7}{2} \\
\epsilon - \frac{7}{2} & \left(\epsilon -\frac{7}{2} \right)^2 +  7 
\end{bmatrix}.
\label{app:highT_thermal_A}
\end{align}

\begin{figure}
\includegraphics[scale=0.55]{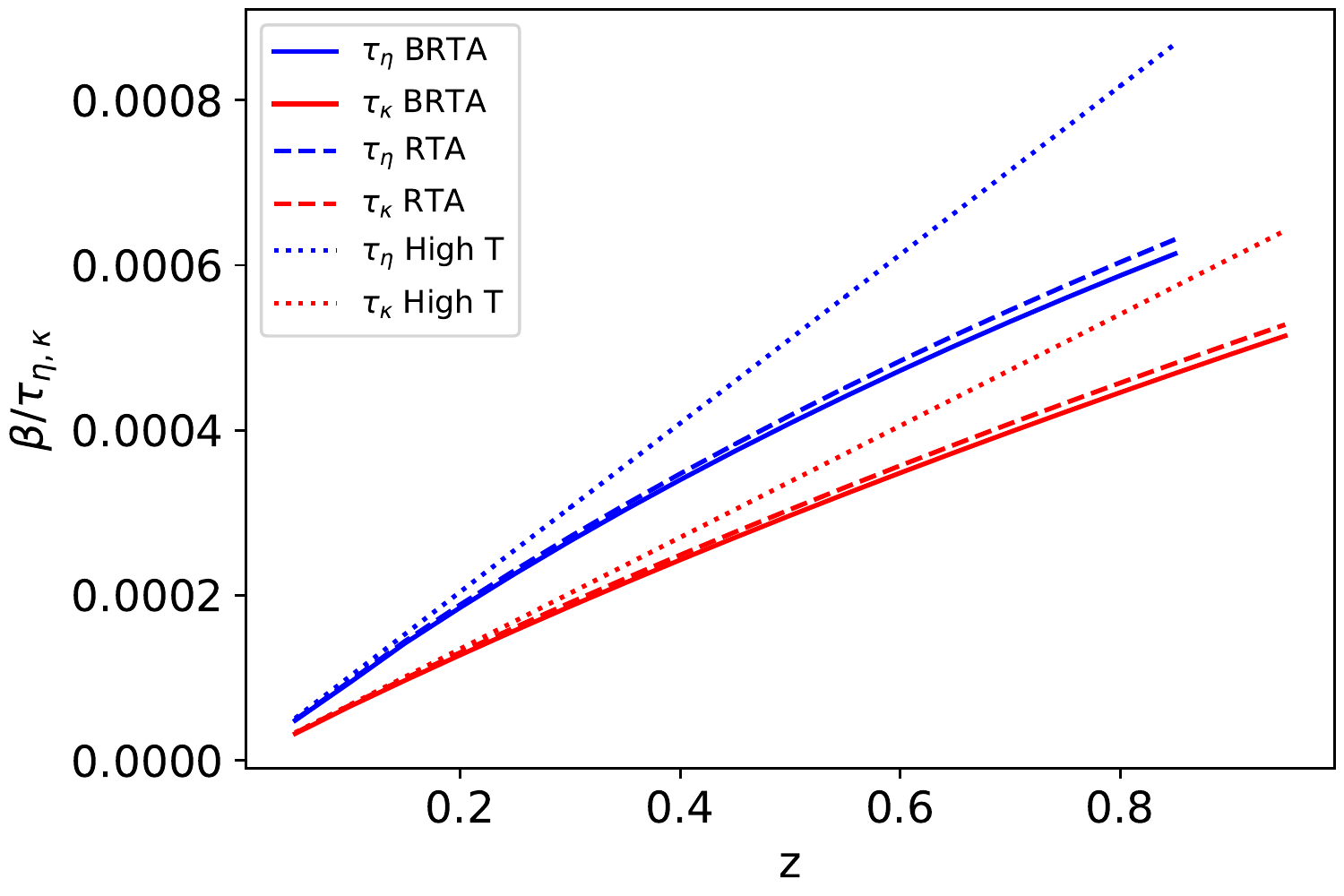}
\caption{Comparing the predictions of the scattering rates, $\tau_{\eta,\kappa}$ from the relaxation time approximation (RTA),  beyond the relaxation time approximation (BRTA), and the high temperature approximation for the BRTA. For this simulation, $\beta E_b = 0$ and $\beta^{1/2} R = 100$. The scattering times in the BRTA are larger than the RTA values, which are smaller than the high-temperature results. This is consistent with the basic structure of the BRTA.} 
\label{fig:brta_to_rta}
\end{figure}

There are several important properties of the matrices $\mathcal{A}^{\eta,\kappa}$. Consider the relaxation time approximation (RTA). In the RTA we only retain the first mode in $\mathcal{A}^{\eta,\kappa}$. This approximation gives the following expression for the scattering times:
\begin{align}
\tau_{\eta}^{(1)} &= \frac{\beta}{\mathcal{A}^{\eta}_{1,1}}, &
\tau_{\kappa}^{(1)} &= \frac{\beta}{\mathcal{A}^{\kappa}_{1,1}}.
\end{align}
Beyond the relaxation time approximation (BRTA), we keep both basis modes, and evaluate the whole matrix, $\mathcal{A}^{\kappa,\eta}$, in order to determine the scattering times. In general one can write the scattering times in the BRTA approximation in terms of the RTA approximation:
\begin{align}
\frac{\tau^{(2)}_{\eta}}{\tau^{(1)}_{\eta}} &= \frac{1}{1-A} & \frac{\tau^{(2)}_{\kappa}}{\tau^{(1)}_{\kappa}} &= \frac{1}{1-B}
\end{align}
where $\tau_{\eta,\kappa}^{(2)}$ are the scattering times in the BRTA approximation, and the constants, $A$ and $B$, are defined as:
\begin{align}
A &=  \frac{\left(\mathcal{A}^{\eta}_{1,2}\right)^2}{\mathcal{A}^{\eta}_{1,1} \mathcal{A}^{\eta}_{2,2}}, & B &=  \frac{\left(\mathcal{A}^{\kappa}_{1,2}\right)^2}{\mathcal{A}^{\kappa}_{1,1} \mathcal{A}^{\kappa}_{2,2}}.
\end{align}
In general the constants $A$ and $B$ are positive because the eigenvalues of $\mathcal A^{\eta,\kappa}$ express the (positive) decay rates of different perturbations. Hence the BRTA prediction for the scattering times will be larger than the RTA prediction; this is in line with the variational formulation that each truncated basis provides a lower bound on the true transport time $\tau$ \cite{Bruun07, Frank20}. In Fig.~\ref{fig:brta_to_rta} we show the scattering rates calculated in both the RTA and BRTA including finite-temperature effects. At high temperatures the results for the BRTA are very close to the RTA, making the RTA a good approximation for the scattering times. As one lowers the temperature, i.e., increases $z$, the difference between the RTA and BRTA becomes more pronounced.

In the high-temperature limit we find $B>A$ from the specific structure of the $\mathcal A^{\eta,\kappa}$ matrices in Eqs.~\eqref{app:highT_shear_A}--\eqref{app:highT_shear_A}, which implies that $\tau^{(2)}_{\kappa}/\tau^{(1)}_{\kappa} > \tau^{(2)}_{\eta}/\tau^{(1)}_{\eta}$. This translates into the following inequality for the Prandtl number calculated in the BRTA, $\Pr^{(2)}$:
\begin{equation}
\Pr^{(2)} = \frac{\tau_{\eta}^{(2)}}{\tau_{\kappa}^{(2)}} = \Pr^{(1)} \frac{1-B}{1-A} \leq \Pr^{(1)},
\end{equation}
where $\Pr^{(1)}$ is the Prandtl number calculated in the RTA.

The BRTA matches the RTA, $\Pr^{(2)} = \Pr^{(1)}$, when $A=B=0$. In general this occurs when the off-diagonal matrix elements vanish, $\mathcal{A}^{\eta,\kappa}_{1,2} = 0$. In the high-temperature limit, $\mathcal{A}^{\eta,\kappa}_{1,2} = 0$ when
\begin{equation}
0 = \int_0^\infty d\epsilon  \frac{\epsilon^5 \left(\epsilon-\frac{7}{2}\right)e^{-\epsilon}}{\beta R^2(\epsilon-\beta E_b)^2 +\epsilon^3}.
\label{eq:condition}
\end{equation}
Given $\beta R^2 \gg 1$, the approximate solutions for Eq.~(\ref{eq:condition}) are $\beta E_b \approx 0, 7/2$. Numerical evaluation of Eq.~(\ref{eq:condition}) gives $\beta E_b \approx 0.186, 3.447$, which is consistent with the red dashed lines in Fig.~\ref{fig:BEC_BCS_Crossover}.

Besides these points where $\Pr^{(2)} = \Pr^{(1)}$, it is important to note the behaviour of the Prandtl number in the strongly interacting (``res.'') limit $\beta^{3/2}v^{-1} = 0$ and in the weakly interacting (``w.i.'') limit $\beta^{3/2}v^{-1} \gg 1$, respectively. A direct evaluation of the scattering times in these two limits yields
\begin{align*}
\left. \tau_{\eta}^{(1)} \right|_\text{res.} &= \frac{\pi}{\sqrt{2}Tz} \frac{25}{576} \frac{R^2}{mT}, & \left. \tau_{\kappa} ^{(1)}\right|_\text{res.} &= \frac{3}{2}\left.\tau_{\eta}^{(1)}\right|_\text{res.} \\
\left. \tau_{\eta}^{(1)} \right|_\text{w.i.} &= \frac{\pi}{\sqrt{2}Tz} \frac{5}{2304} \frac{1}{v^2 (mT)^3}, & \left. \tau_{\kappa}^{(1)} \right|_\text{w.i.} &= \frac{3}{2}\left.\tau_{\eta}^{(1)} \right|_\text{w.i.}
\end{align*}
within the RTA, while in the BRTA
\begin{align*}
\left. \tau_{\eta}^{(2)} \right|_\text{res.} &= \left(1+\frac{3}{202}\right) \left.\tau_{\eta}^{(1)} \right|_\text{res.} & \left. \tau_{\kappa}^{(2)} \right|_\text{res.} &= \left(\frac{3}{2}+ \frac{3}{88}\right) \left.\tau_{\eta}^{(1)} \right|_\text{res.} \\
\left. \tau_{\eta}^{(2)} \right|_\text{w.i.} &= \left(1+\frac{75}{226}\right) \left.\tau_{\eta}^{(1)} \right|_\text{w.i.} & \left. \tau_{\kappa}^{(2)} \right|_\text{w.i.} &= \left(\frac{3}{2} + \frac{75}{104} \right) \left.\tau_{\eta}^{(1)} \right|_\text{w.i.}.
\end{align*}
This yields the scattering ratios in the strongly and weakly interacting limits that are quoted in the main text in Eq.~\eqref{eq:scattering_ratios}.

\section{Angular Averages}
\label{appendix:angular_averages}

We report some general formulas for angular averages of functions $f(x = {\bf \hat{Q} \cdot \hat{p}})$ over a variable number of unit vectors, $\hat{\bf p}$. The calculations are straight forward, but become rather tedious. We restrict ourselves to an even number of unit vectors:
\begin{align}
\langle f(x) \rangle_{\hat{\bf p}}= \int\frac{d\Omega_{\bf p}}{4\pi} f(x)  = \int_{-1}^{1} \frac{dx}{2} f(x),
\label{app:avg_0}
\end{align}
\begin{align}
\langle \hat{\bf p}_i \hat{\bf p}_j  f(x) \rangle_{\hat{\bf p}} &= \int\frac{d\Omega_{\bf p}}{4\pi} \hat{\bf p}_i\hat{\bf p}_j f(x)  \nonumber \\
&= \int_{-1}^1 \frac{dx}{2} f(x) \left[\frac{1-x^2}{2}\delta_{i,j} + \hat{\bf Q}_i\hat{\bf Q}_j P_2(x) \right],
\label{app:avg_2}
\end{align}
\begin{align}
\langle \hat{\bf p}_i & \hat{\bf p}_j \hat{\bf p}_k \hat{\bf p}_l   f(x) \rangle_{\hat{\bf p}} =\int\frac{d\Omega_{\bf p}}{4\pi}  \hat{\bf p}_i\hat{\bf p}_j\hat{\bf p}_k\hat{\bf p}_l f(x) =\nonumber \\
& \int_{-1}^1 \frac{dx}{2} f(x)  \left[\frac{(1-x^2)^2}{8}\left(\delta_{i,j}\delta_{k,l} + \text{2 permutations}\right) \right. \nonumber \\
& +\frac{-5 x^4 + 6 x^2 -1}{8} \left(\hat{\bf Q}_i\hat{\bf Q}_j \delta_{k,l} + \delta_{i,j} \hat{\bf Q}_k\hat{\bf Q}_l + \text{2 perms.}\right) \nonumber \\
&\left. + \hat{\bf Q}_i\hat{\bf Q}_j\hat{\bf Q}_k\hat{\bf Q}_lP_4(x) \right],
\label{app:avg_4}
\end{align}
\begin{align}
\langle \hat{\bf p}_i &\hat{\bf p}_j \hat{\bf p}_k \hat{\bf p}_l \hat{\bf p}_r \hat{\bf p}_s  f(x) \rangle_{\hat{\bf p}} =\int\frac{d\Omega_{\bf p}}{4\pi} \hat{\bf p}_i\hat{\bf p}_j\hat{\bf p}_k\hat{\bf p}_l\hat{\bf p}_r\hat{\bf p}_s f(x) =\nonumber \\
& \int_{-1}^1 \frac{dx}{2} f(x)  \left[\frac{(1-x^2)^3}{48}\left(\delta_{i,j}\delta_{k,l}\delta_{r,s} + \text{14 perms.}\right) \right. \nonumber \\
& +\frac{7x^6 -15 x^4 + 9 x^2 -1}{48} \nonumber \\
&\left(\hat{\bf Q}_i\hat{\bf Q}_j \delta_{k,l} \delta_{r,s} + \delta_{i,j} \hat{\bf Q}_k\hat{\bf Q}_l \delta_{r,s} + \delta_{i,j} \delta_{k,l} \hat{\bf Q}_r\hat{\bf Q}_s\right.\nonumber \\
&\left. + \text{14 perms.}\right) \nonumber \\
&+ \frac{-21 x^6 + 35 x^4 - 15x^2 +1}{48} \nonumber \\
&\left(\hat{\bf Q}_i\hat{\bf Q}_j \hat{\bf Q}_k\hat{\bf Q}_l \delta_{r,s} + \hat{\bf Q}_i\hat{\bf Q}_j \delta_{k,l} \hat{\bf Q}_r\hat{\bf Q}_s \right. & \nonumber \\
&\left. + \delta_{i,j} \hat{\bf Q}_k\hat{\bf Q}_l \hat{\bf Q}_r\hat{\bf Q}_s + \text{14 perms.} \right) \nonumber \\
&\left. + \hat{\bf Q}_i\hat{\bf Q}_j\hat{\bf Q}_k\hat{\bf Q}_l\hat{\bf Q}_r\hat{\bf Q}_sP_6(x) \right].
\label{app:avg_6}
\end{align}
In these equations $P_n(x)$ is the Legendre polynomial. We note that the only term proportional to $P_0(x)$ is the term that is independent of the center-of-mass momentum, ${\bf Q}$. Thus if $f(x)$ does not depend on $x$, only the first lines of Eqs.~(\ref{app:avg_0}-\ref{app:avg_6}) are nonzero, reducing to the standard formulae for averages of unit vectors.

\section{The Bose-Fermi Model}
\label{appendix:Bose_Fermi}
In this appendix we review the thermodynamics of the Bose-Fermi model, which describes a non-interacting mixture of bosons with mass $2m$ and chemical potential $2\mu-E_b \theta(-E_b)$, and fermions of mass $m$ and chemical potential $\mu$. The pressure of the system is then given by:
\begin{align}
P = \frac{1}{(2\pi \beta)^{3/2}\beta} &\left[ -\Li_{5/2}(-z) \right.  \nonumber \\
&\left. + 3 \times 2^{3/2} \Li_{5/2}\left(z^2 e^{-\beta E_b \theta(-E_b)}\right)\right],
\label{app:bose_fermi_pressure}
\end{align}
where $\Li_a(z)$ is the polylogarithm function. In Eq.~(\ref{app:bose_fermi_pressure}), the first term is related to the noninteracting Fermi gas, while the second term describes the noninteracting bosons.  The factor of $3$ originates from the $2\ell + 1$ degeneracy of the p-wave coupling, i.e., there are $3$ bound states corresponding to coupling in the $m_{\ell} = -1,0,1$ channels, where $m_{\ell}$ is the azimuthal quantum number.

In this model the density is given by
\begin{align}
n = \frac{1}{(2\pi\beta)^{3/2}}&\left[-\Li_{3/2}(-z) \right. \nonumber \\
&\left. + 6\times2^{3/2} \Li_{3/2}\left(z^2 e^{-\beta E_b \theta(-E_b)}\right) \right].
\label{app:bose_fermi_density}
\end{align}

\noindent From Eqs.~(\ref{app:bose_fermi_pressure}-\ref{app:bose_fermi_density}) one can calculate the specific heat at constant pressure per unit volume and the density for arbitrary temperature in the normal phase \cite{Yao18}. In this article we primarily focus on the limit where $|\beta E_b|$ is small, hence we ignore the bound state energy and its correction to the bosonic chemical potential.


\begin{thebibliography}{References}
\bibitem{T_matrix_Note} In principle the two-body T-matrix can be a rank-2 tensor of the form: $\langle {\bf Q}/2 \pm {\bf p} | \mathcal{T} | {\bf Q}/2 \pm {\bf q} \rangle = p_i q_i T_{i,j}({\bf Q}, E)$, where $T_{i,j}({\bf Q},E) = \delta_{i,j} T^{(0)}({\bf Q},E) + Q_i Q_j T^{(1)}({\bf Q},E)$. However, $T^{(1)}$ vanishes like $\Lambda^{-3}$, where the ultraviolet cutoff $\Lambda$ is sent to infinity.
\end{thebibliography}
\end{document}